\RequirePackage{lineno}
\documentclass[superscriptaddress,amsmath,amssymb,nofootinbib]{revtex4}

\usepackage{graphicx}
\usepackage{dcolumn}
\usepackage{bm}
\usepackage{ulem}
\usepackage{xfrac}

\usepackage[printwatermark]{xwatermark}
\usepackage{xcolor}

\usepackage{graphicx,amsmath,amssymb}
\usepackage{epstopdf}
\usepackage{lineno}
\usepackage[colorlinks=true, urlcolor=blue, linkcolor=blue, citecolor=blue]{hyperref}

\begin{document}

\title{An explanation for dark matter and dark energy consistent
with the Standard Model of particle physics and General Relativity}

\author{ Alexandre Deur}

\affiliation{University of Virginia, Charlottesville, VA 22904. USA\\
\email{deurpam@jlab.org} 
}



\begin{abstract}

Analyses of internal galaxy and cluster dynamics typically employ Newton's law of gravity, which neglects the 
field self-interaction effects of General Relativity. This may be why dark 
matter seems necessary. The Universe evolution, on the other hand, is treated with the full theory, General Relativity. 
However, the approximations of isotropy and homogeneity, normally used to derive and solve the Universe evolution equations,
effectively suppress General Relativity's field self-interaction effects and this may introduce the need for dark energy. 
Calculations have shown that field self-interaction increases the binding of matter inside massive systems,
which may account for galaxy and cluster dynamics without invoking dark matter. In turn, 
energy conservation dictates that the increased binding must be balanced by an effectively decreased 
gravitational interaction outside the massive system. In this article, such suppression is estimated and its 
consequence for the Universe's evolution is discussed. Observations 
are reproduced without need for dark energy. 

\end{abstract}

\maketitle

\section{Introduction}
For the last 20 years, observations have shown that the Universe's
expansion is presently accelerating. The first solid indication came from
measurements of the apparent magnitude of supernovae~\cite{large-z 98 data,large-z 99 data}. 
The leading explanations for the origin of the acceleration are either a non-zero
cosmological constant $\Lambda$, or exotic fields~\cite{review dark energy}.
This article investigates another possibility which does not require
$\Lambda \neq 0$, exotic fields, or a modification of General Relativity 
(GR). This alternative is a direct consequence of a mechanism that can explain the missing 
mass problem in galaxies and galaxy clusters
without requiring dark matter nor modifying gravity/dynamical laws~\cite{Deur DM-PLB,Deur DM-EPJC}. 
The phenomenology stems from GR's field self-interaction, which causes
GR's non-linear behavior\footnote{"Self-interaction" is used rather than the less specific "non-linear" denomination: 
non-linearities in GR or QCD arise from field self-interaction. 
In contrast, pure-field QED is a linear theory. Non-linearities appear in QED (e.g. photon-photon 
scattering) once matter is introduced. To distinguish between these two cases, "self-interaction" is used.}.
The consequences of such field self-interaction are well-studied in Quantum Chromodynamics 
(QCD) which Lagrangian has a similar structure to that of GR. 

GR's Lagrangian density is: 
\begin{eqnarray}
\mathcal{L}_{\mathrm{GR}}=\frac{1}{16\pi G}\sqrt{\mathrm{det}(g_{\mu\nu})}\, g_{\mu\nu}R^{\mu\nu}, 
\label{eq:Einstein-Hilbert Lagrangian}
\end{eqnarray}
where $G$ is the Newton constant, $g_{\mu\nu}$ the metric and $R_{\mu\nu}$ the Ricci
tensor. The deviation of $g_{\mu\nu}$ from a constant reference metric $\eta_{\mu\nu}$ 
defines the gravity field, $\psi_{\mu\nu} = g_{\mu\nu} - \eta_{\mu\nu}$. 
Expanding in $\psi_{\mu\nu}$ and rescaling\footnote{The magnitude of the gravity field $\psi_{\mu\nu}$ 
being proportional to the quantity of matter, 
$\psi^2 \propto M$, the rescaled field  $\sqrt{M}\varphi_{\mu\nu} = \psi_{\mu\nu}$ is the field originating from a unitary mass.
This notation emphasizes that effectively, self-interaction terms couple as $\sqrt{GM}$. The rescaling does not affect
the results since it amounts to rescaling $\mathcal{L}_{GR}$ by $1/M$ and
for classical systems, $\mathcal{L}$ can be rescaled without physical effects.
}
the field as $\varphi_{\mu\nu} = \psi_{\mu\nu}/\sqrt{M}$ yields the field Lagrangian~\cite{Zee}:
\begin{eqnarray}
\mathcal{L}_{\mathrm{GR}}\!=\!\left[\partial\varphi\partial\varphi\right]\!+\!\sqrt{16\pi MG}\left[\varphi\partial\varphi\partial\varphi\right]\!+ 
\!16\pi MG\left[\varphi^{2}\partial\varphi\partial\varphi\right]\!+\cdots ,
\label{eq:Polynomial Einstein-Hilber Lagrangian}
\end{eqnarray}
where  $\left[\varphi^{n}\partial\varphi\partial\varphi\right]$
denotes a sum over Lorentz-invariant terms of the form $\varphi^{n}\partial\varphi\partial\varphi$, and 
$M$ is the system mass. 
The $n>0$ terms cause field self-interaction, i.e. the non-linearities that distinguish GR from
Newton's theory. This latter is  given by the $n=0$ term, $\mathcal{L}_{\mathrm{Newton}} = \left[\partial\varphi\partial\varphi\right]$.

QCD's field Lagrangian is:
\begin{eqnarray}
\label{eq:QCD Lagrangian}
\mathcal{L}_{QCD}\!=\!\left[\partial \phi \partial \phi \right]\!+\!\sqrt{\pi \alpha_s}\left[\phi^2 \partial \phi \right]\!+\!
\pi \alpha_s\left[\phi^4\right], 
\end{eqnarray}
with $\phi_\mu^a$ the gluonic field and $\alpha_s$ the QCD coupling. In the bracket terms,
contractions of the color charge indices $a$ are understood in addition to the sums over Lorentz-invariant terms. 
As in Eq.~(\ref{eq:Polynomial Einstein-Hilber Lagrangian}), field self-interaction arises from 
the terms beside $\left[\partial \phi \partial \phi \right]$. 
Those stem from the color charges carried by the gluonic field.
Likewise, GR's self-interaction originates from its field's energy-momentum, the tensor-charge to which gravity couples.

In QCD, self-interaction effects are conspicuous because $\alpha_{s}$
is large, typically $\simeq0.1$ at the transition between QCD's weak 
and strong regimes~\cite{Deur:2016tte}. A crucial consequence is an increased 
binding of quarks, which leads to their confinement.
In GR, self-interaction becomes important for $\sqrt{GM/L}$ large
enough ($L$ is the system characteristic scale), typically for 
$\sqrt{GM/L}\gtrsim10^{-3}$ as discussed in Ref.~\cite{Deur DM-EPJC}
or exemplified by the Hulse-Taylor binary pulsar~\cite{Hulse:1974eb}, the first system
in which GR was experimentally tested in its strong regime, which has $\sqrt{G M/L} = 10^{-3}$.
As in the case of QCD, self-interaction increases the binding
compared to  Newton's theory. Since the latter is used to treat the internal
dynamics of galaxies or galaxy clusters, its neglect of self-interaction may contribute to --or even create-- the
missing mass problem~\cite{Deur DM-PLB,Deur DM-EPJC,Deur:2013baa}.
In Ref.~\cite{Deur DM-PLB} a non-perturbative numerical calculation based  on Eq.~(\ref{eq:Polynomial Einstein-Hilber Lagrangian})
is applied  in the static limit to spiral galaxies and clusters. 
A non-perturbative formalism (lattice technique) 
--rather than a perturbative one such as the post-newtonian formalism--
was chosen because in QCD, confinement
is an entirely non-perturbative phenomenon, unexplainable within a perturbative approach.
The results of Refs.~\cite{Deur DM-PLB,Deur DM-EPJC} indicate that self-interaction
increases sufficiently the gravitational binding of large massive systems such that no dark
matter nor ad-hoc gravity/dynamical law modifications are needed to account for the galaxy
missing mass problem. Self-interaction also explains
galaxy cluster dynamics and the Bullet cluster observation~\cite{bullet cluster}.
Finally, the Tully-Fisher relation~\cite{Tully:1977fu}, an important observation difficult to explain in the dark matter context,
was shown in Ref.~\cite{Deur DM-PLB} to be the GR analog to QCD's Regge trajectories~\cite{the:Regge}. 
Accounting for self-interaction automatically yields flat rotation curves 
for disk galaxies when those are modeled as homogeneous disks of baryonic matter
with exponentially decreasing density profiles, which is a good approximation of the observations.
In contrast, dark matter halo profiles must be specifically tuned for each galaxy to make its rotation curve flat.

Besides quark confinement, the other principal feature 
of QCD is a dearth of strong interaction outside of hadrons (i.e. quark bound states) because color confinement
keeps the (colored) gluonic field in the hadron. 
As shown in numerical lattice  calculations~\cite{Bali:1994de}, the field lines 
--which for a free-field spread isotropically from the source to infinite distances-- are for a self-interacting field 
rearranged in a finite volume roughly contained between the quarks: the collapsed field lines between two 
quarks form an approximately one dimensional ``flux-tube'' in which flux lines do not spread. 
Their density, i.e. the force acting between quarks, is hence constant with the quark separation $r$.
While this confined field produces a binding energy stronger than in the free-field case, 
such field concentration inside the hadron causes a field depletion outside. This conforms to
energy conservation: compared to the free-field case, the increased binding energy in 
the hadron 
is compensated by a near absence of potential energy outside the hadron 
since the field lines 
have been pulled-in due to self-interaction.
Increases of binding energy have also been calculated, with the same numerical lattice technique, 
for gravity and massive structures~\cite{Deur DM-PLB,Deur DM-EPJC}. This increased binding must, by 
energy conservation, weaken the action of gravity at larger scale.
This can then be mistaken for a repulsion, i.e. dark
energy. Specifically, the Friedmann equation for an isotropic and homogeneous Universe
is (for a matter-dominated flat Universe) $H^{2}=8\pi G\rho/3$,
with $H$ the Hubble parameter and $\rho$ the density. As massive structures coalesce, 
gravity is effectively suppressed at scales larger that these structures. This weakening with time results in
a larger than expected value of $H$ at early times, as seen by the
observations suggesting the existence of dark energy. 


An important point for the present article is that the morphology of the massive structures in which
gravity may be trapped determines how effective the trapping is: the less isotropic and homogeneous a system is, 
the larger the trapping is.
For example, this implies a correlation between the missing mass
of elliptical galaxies and their ellipticities. The correlation was predicted in~\cite{Deur DM-PLB} and
subsequently verified in~\cite{Deur:2013baa}. 
The role of the system spacial symmetry is also supported by the relation $J = \epsilon M^\gamma$ 
describing both the galactic Tully-Fisher observation and the hadronic Regge trajectories.\footnote{The 
Tully-Fisher relation is usually expressed as $L \propto V^x $, with $L$ the absolute luminosity
of the galaxy (proportional to its visible mass $M$), $V$ the rotation speed and $x=3.9\pm0.2$. 
Since $MV \propto J$ with $J$ the disk angular momentum, the Tully-Fisher relation can then be 
re-expressed as $J \propto M^{1.26\pm0.07}$, of the same form as Regge trajectories $J \propto M^2$.
Regge trajectories stem from the increase of the quark binding energy 
necessary to compensate for the increased centrifugal force at higher angular momentum $J$.
The potential determining the binding energy (essentially the mass for light hadrons) is  proportional to 
$r$, which yields $J \propto M^2$.
A similar picture holds for the Tully-Fisher relation in the self-interaction framework. The difference is 
the shape of the system in which the force is confined, 
{\it {viz}} the 2-dimensional galaxy disk  rather than the 1-dimensional
flux tube. In two dimensions, flux lines density, i.e. force, falls as $1/r$, yielding in a $\ln(r)$ potential not as steep as 
the 1-dimensional potential proportional to $r$.} 
Here $J$ is the angular momentum, $M$ the system mass, 
and $\epsilon$ a constant depending on the type of galaxy or hadron  family 
considered.
Inside a less symmetric system, the force is more enhanced and $\gamma$ is larger than that of a 
more symmetric system, as observed: Regge trajectories apply to hadrons (flux tubes of 1-dimension) and 
have $\gamma=2$, 
while for the Tully-Fisher relation which applies to disk galaxies (2-dimensional systems), $\gamma = 1.26\pm0.07$.

The possible effects of the Universe's inhomogeneity have been discussed in the past 
to explain cosmological observations without requiring dark energy~\cite{Mannheim:2005bfa}. In particular,
the possible importance of backreactions, of same origin as field self-interaction, has been pointed out~\cite{Buchert:2007ik}.
The calculations carried out so far are typically perturbative. Thus they are blind to the non-perturbative
phenomena that are critical in the analogous QCD phenomenology. 
Previous non-perturbative attempts have been inconclusive~\cite{Buchert:2013qma}.
The present approach, while remaining within GR's description of the Universe evolution, 
see Section~\ref{Universe evolution},  folds the effects of inhomogeneities  into a generic function $D_M$ 
that expresses the large distance consequences of the non-perturbative effects, and
which functional form is modeled from general 
considerations, see Section~\ref{Construction of D}.
That this approach differs from others using backreactions or inhomogeneity
is illustrated by the identification of an explicit mechanism (field trapping) that is not perturbative, 
and by the direct connection between 
dark energy and dark matter that this work exposes.

In summary, traditional analyses of internal galaxy  or cluster dynamics employ Newton's gravity that
neglects the self-interaction terms in Eq.~(\ref{eq:Polynomial Einstein-Hilber Lagrangian}),
and this may explain the need for dark matter~\cite{Deur DM-PLB,Deur DM-EPJC}.
Traditional analyses of Universe evolution do use GR, but under the approximations of
 isotropy and homogeneity, which suppress the effects of the self-interaction 
 terms~\cite{Deur DM-PLB,Deur:2013baa}, and would by definition disregard any local phenomenon
 that could affect gravity's field, such as field trapping. 
The weakening of gravity at large distance due to these terms is thus neglected, which may
be why dark energy seems necessary. This was conjectured in Ref.~\cite{Deur DM-PLB} 
and the present article investigates this possibility.

\section{Field depletion outside massive structures}\label{depletion}
As just discussed, energy conservation implies that
the increased binding energy in massive non-isotropic systems,
e.g. galaxies or galaxy clusters, should decrease
gravity's influence outside these systems.\footnote{In GR, total energy is not necessarily conserved 
since its definition may excludes gravitational
energy. However, here (and in the previous work of Ref.~\cite{Deur DM-EPJC})
gravitational energy
is included. In any cases, regardless of its definition,
energy is conserved in static cases or for asymptotically flat space-time,
which are the cases treated here and in~\cite{Deur DM-EPJC}. 
} The consequence
on the Universe's dynamics can be folded in a depletion factor $D$. 
We will show in Section~\ref{Universe evolution} that such factor naturally 
appears in the Universe's evolution equations once the approximations 
that the Universe is isotropic and homogeneous are lifted.
Before this, 
 it is useful to first get an idea of its form and magnitude. Its more thorough determination 
 is given in Section~~\ref{Construction of D}.
 
When self-interaction effects are small (low mass) or suppressed (symmetric system) 
$D\simeq1$ and the traditional treatment of gravity applies. 
When gravity's field is trapped in a massive system, gravity is suppressed outside the system. This is
accounted for by having $D<1$, with $D=0$ if the field is fully trapped. 
Since the spacial distributions of matter, radiation, and dark energy differ, 
separate $D$ factors must be considered for these quantities, $D_{M}$, $D_{R}$ 
and $D_{\Lambda}$, respectively. Electromagnetic
radiation does not clump and couples weakly to gravity, so $D_{R}\simeq1$.
Presumably, $D_{\Lambda}=1$ for the same reason but in any case, we assume 
$\Lambda=0$ throughout this article. Since self-interaction effects disappear for
homogeneous isotropic systems, $D_{M}\simeq1$ for the early Universe. Then
$D_{M}$ decreases as structure formation renders the  Universe less homogeneous. 
Thus, $D_{M}$ depends on time, i.e. on the redshift $z$, and this dependence 
is driven by large structure formation. In particular, significant field  trapping occurs,
i.e. the transition from $D_{M}(z)\simeq1$ to $D_{M}(z)<1$,
when galaxies formed and became massive enough so that the 
$\sqrt{GM}$ coupling in Eq.~(\ref{eq:Polynomial Einstein-Hilber Lagrangian}) enables self-interactions.
This happens in the range $2\lesssim z\lesssim10$ since typically, present ($z=0$) galaxies have
$\sqrt{GM/L} \approx 10^{-3}$ and a large structure mass increases as $(1+z)^{-1}$. ($L $ grows slower
so we ignore its $z$-dependence in the assessment.) $D_{M}(z)$ then continues to change as
groups and clusters form. $D_{M}(z)$ may not always decrease with time
even if the structures' masses increase, since trapping also depends
on the homogeneity and symmetry of the structures. 
For instance, some galaxies had filament shapes for $z\gtrsim2$, which favors field trapping and thus small $D_M$. They then
grew to disks or ellipsoids, i.e. more symmetric morphologies for which field self-interaction effects tend to cancel out. 
Likewise, the elliptical/disk galaxy ratio is continuously 
increasing~\cite{the:E/S ratio evolution3,the:E/S ratio evolution4,the:E/S ratio evolution5}.
This implies that, all other things being equal (e.g. ignoring other rearrangements such as in clusters), 
$D_{M}$ may rise at small $z$. 

\section{Accounting for field depletion in evolution of the Universe}\label{Universe evolution}
Once structures have enough mass so that GR's self-interaction cannot be neglected anymore, 
field trapping diminishes the effect of gravity at scales larger than the structures.
One can thus presume that the effect can be embodied by a function $D(z)$ factoring the gravity magnitude $G$. 
We demonstrate it in this section by tracking the terms that
disappear from the Universe evolution equation under the hypotheses of homogeneity and isotropy, 
and by identifying these terms with the effect of field trapping.\footnote{
The same method is used in QCD: when symmetry assumptions are lifted, 
new structures terms appear e.g. in cross-section expressions. These terms are then parameterized
using measurements or calculated non-perturbatively e.g. with lattice methods.}    
First, we recall the traditional evolution equation obtained assuming homogeneity and isotropy.

\subsection{Evolution equation for an homogeneous, isotropic Universe }\label{Evol. eqs. without D}

The Universe evolution equation is derived using the Einstein field equation:
\begin{eqnarray}
R_{\mu\nu}=-8\pi GS_{\mu\nu},
\label{eq: Einstein Field eq.}
\end{eqnarray}
with $S_{\mu\nu}$ the energy-momentum tensor.  Assuming an 
homogeneous and isotropic Universe reduces $R_{\mu\nu}$ and $S_{\mu\nu}$ to diagonal tensors. 
Eq.~(\ref{eq: Einstein Field eq.}) then yields:
 \begin{eqnarray}
R_{00}=\frac{3\overset{\mathbf {..}}{a}}{a}, \label{standard evol eqs.1}\\
R_{ij}=-\big[2K+2\overset{\mathbf .}{a}^{2}+a{\overset{\mathbf {..}}{a}}\big]g_{ij},\\
R_{0i}=0,\\
S_{ij}=\frac{1}{2}(\rho-p)a^2g_{ij},
\label{standard evol eqs.2}
\end{eqnarray}
where  $a$ is the Robertson-Walker scale factor, $K$ the space curvature sign,  $\rho$ the density,  
$p$ the pressure, and latin indices denote spacial components only.
Combining Eqs.~(\ref{standard evol eqs.1}-\ref{standard evol eqs.2}) yields the traditional Friedmann equation:
\begin{eqnarray}
\overset{\mathbf .}{a}^{2}+K=8\pi G \rho a^{2}/3. \label{eq: original Friedmann}
\end{eqnarray}

\subsection{Evolution equation for an inhomogeneous, anisotropic Universe }\label{Evol. eqs. with D}
Structure formation causes spatial inhomogeneities and, once those are massive enough, field trapping is
induced. Terms in $R_{\mu\nu}$ and $S_{\mu\nu}$ that vanish under the approximations of isotropy and 
homogeneity now appear with the formation of structures. We show here that within GR's formalism, 
these terms can be regrouped in an overall term $\mathbf D(z)$ factoring the right hand side of 
Eq.~(\ref{eq: original Friedmann}).

If the assumptions of isotropy and homogeneity are lifted, new terms, including off-diagonal ones, appear 
in $R_{\mu \nu}$ and $S_{ij}$. Eqs.~(\ref{standard evol eqs.1}-\ref{standard evol eqs.2}) then change to:
 \begin{eqnarray}
R_{00}(1+\alpha) & = &\frac{3\overset{\mathbf {..}}{a}}{a}, \label{R00} \\
R_{ik}(\delta^k_j+\beta^k_j) & = &-\big[2K+2\overset{\mathbf .}{a}^{2}+a{\overset{\mathbf {..}}{a}}\big]g_{ij}, \label{Rij} \\
R_{0i} & =& \gamma_i, \label{R0i} \\
S_{ik} & = & \frac{1}{2}(\rho-p)a^2g_{ij}(\delta^j_k+\theta^j_k), \label{Sij}
 \end{eqnarray}
where $\delta_{ij}$ is the Kronecker delta, and $\alpha$, $\beta_{ij}$, $\gamma_{i}$ and $\theta_{ij}$ 
are functions representing the anisotropic components of $R_{\mu \nu}$ and $S_{ij}$, 
i.e. the components vanishing when isotropy and the Robertson-Walker metric are assumed. 
Combining Eqs.~(\ref{eq: Einstein Field eq.}) and (\ref{R00}) yields:
\begin{eqnarray}
\frac{3\overset{\mathbf {..}}{a}}{a}  = -4\pi G(\rho+3p)(1+\alpha).
\label{eq. gen. E+RW 1}
 \end{eqnarray}
Eqs.~(\ref{eq: Einstein Field eq.}) and (\ref{Rij}) together bring:
\begin{eqnarray}
-8\pi G S_{ik} (\delta^k_j+\beta^k_j)=-\big[2K+2\overset{\mathbf .}{a}^{2}+a{\overset{\mathbf {..}}{a}}\big]g_{ij}.
\label{eq. intermediate}
 \end{eqnarray}
Combining Eqs.~(\ref{Sij}) and (\ref{eq. intermediate}) gives:
\begin{eqnarray}
4\pi G(\rho-p)(1+\omega)=\bigg[\frac{2K}{a^2}+\frac{2\overset{\mathbf .}{a}^{2}}{a^2}+\frac{\overset{\mathbf {..}}{a}}{a}\bigg],
\label{eq. gen. E+RW 2}
 \end{eqnarray}
where $\omega \equiv g_{il}(\beta^l_j+\theta^l_j+\theta^l_k\beta^k_j)(g^{-1})^{ij}$ 
i.e. $\omega$ is an average of the anisotropy factors. 
Defining $D(z) \equiv \big[ (1+\frac{3\omega+\alpha}{4})+\frac{3p}{4\rho}(\alpha-\omega) \big] $,
Eqs.~(\ref{eq. gen. E+RW 1}) and (\ref{eq. gen. E+RW 2}) yield:
%
%
\begin{eqnarray}
\overset{\mathbf .}{a}^{2}+K=8\pi G D(z) \rho a^{2} /3. \label{eq:Friedmann}
\end{eqnarray}
We assumed here that the effect of field trapping are represented by $D(z)$ which we thus identify to 
the depletion function.\footnote{Most generally, the inhomogeneity/anisotropy term $D(z)$ 
may contain effects other than field-trapping (like e.g. a pure QCD 
calculation of a structure function would not include electromagnetic effects on hadron structure). 
We neglect this possibility. 
}
This fulfills the expectation that $D(z)$ factors $G$ and
that $D(z) \to 0$ for an homogeneous isotropic Universe. 
 
For simplicity  we have not distinguished so far between non-relativistic matter and radiation/relativistic matter. 
The anisotropy factors for the latter, $\omega_R$ and $\alpha_R$, are negligible. 
So, in the early Universe, when all content is relativistic, $(\alpha_R - \omega_R) \approx 0$. In the latter Universe, 
 $\rho \gg p$ so $\frac{3p}{4\rho}( \alpha_M - \omega_M) \ll [1+(3\omega_M + \alpha_M)/4]$. 
 ($\omega_M$ and $\alpha_M$ are the anisotropy factors for non-relativistic matter.)
Accounting for this simplifies  the depletion function to:
\begin{eqnarray}
D(z) \equiv \bigg[ \bigg(1+\frac{3\omega+\alpha}{4}\bigg)+\frac{3p}{4\rho}(\alpha-\omega) \bigg] \approx 1+(3\omega+\alpha)/4. 
\end{eqnarray}

That anisotropy factors differ for relativistic and non-relativistic contents and for $\Lambda$ is 
formalized by transforming $D$ and $\rho$ into vectors in Eq.~(\ref{eq:Friedmann}), with 
$\rho \to \boldsymbol{\rho}=(\rho_{M},\rho_{R},\rho_{\Lambda})$ 
and  $D \to \mathbf{D}=\left(D_{M},D_{R},D_{\Lambda}\right)$.
After the matter-radiation equilibrium epoch, $z\ll z_{eq}\simeq3400$,
and assuming $\Lambda = 0$, one has $\mathbf{D}\simeq\left(D_{M},1,1\right)$
and $\boldsymbol{\rho} \simeq(\rho_{M},0,0)$.

The present-time critical density $\boldsymbol{\rho_{c{0}}}$ is defined by setting
$K=0$ in the vector version of Eq.~(\ref{eq:Friedmann}):
\begin{eqnarray}
\boldsymbol{\rho_{c{0}}}\mathbf D(0)\equiv\frac{3H_{0}^{2}}{8\pi G},\label{eq:critical density}
\end{eqnarray}
with 
$H_{0} \equiv \overset{\mathbf .}{a}_{0}/a_{0}$.
The densities of matter, radiation and  $\Lambda$ evolve
as usual:
\begin{eqnarray}
\boldsymbol{\rho} & =\left(\rho_{0 {M}}\left(\frac{a_{0}}{a}\right)^{3},\rho_{0 {R}}\left(\frac{a_{0}}{a}\right)^{4},\rho_{0 {\Lambda}}\right).\label{eq:density 1}
\end{eqnarray}
Defining 
$\Omega_{M}^{*}(z)\equiv\frac{8\pi GD_{M}(z)}{3H_{0}^{2}}\rho_{0 {M}}$,
$\Omega_{R}^{*}\equiv\frac{8\pi GD_{R}}{3H_{0}^{2}}\rho_{0 {R}}$ and
$\Omega_{\Lambda}^{*}\equiv\frac{8\pi GD_{\Lambda}}{3H_{0}^{2}}\rho_{0 {\Lambda}}$,
Eqs.~(\ref{eq:Friedmann}) and (\ref{eq:density 1}) yield:
\begin{eqnarray}
\boldsymbol{\rho} \mathbf D(0) & =\frac{3H_{0}^{2}}{8\pi G}\left[\Omega_{M}^{*}\left(\frac{a_{0}}{a}\right)^{3}+\Omega_{R}^{*}\left(\frac{a_{0}}{a}\right)^{4}+\Omega_{\Lambda}^{*}\right].\label{eq:density 2}
\end{eqnarray}
A ``screened'' density fraction $\Omega^{*}$ has the
form $\Omega^{*}=\Omega D$ where $\Omega$ corresponds
to the traditional definition: $\Omega\equiv\frac{8\pi G}{3H_{0}^{2}}e_{0}$.
The $\Omega^{*}$ are not directly comparable with the mass-energy
census of the Universe. They are relevant to densities assessed from
the Universe dynamical evolution. With the definition of $\Omega^{*}$,
$\boldsymbol{\rho}$ is explicitly independent of $\mathbf D(z)$. Eq.~(\ref{eq:Friedmann})
yields for present time:
\begin{eqnarray}
1 =\left[D_{M}(0)\Omega_{M}+D_{R}\Omega_{R}+D_{\Lambda}\Omega_{\Lambda}\right]-\frac{K}{a_{0}^{2}H_{0}^{2}},
\label{eq:density 3}
\end{eqnarray}
 which leads to $\Omega_{K}\equiv-\frac{K}{a_{0}^{2}H_{0}^{2}}$,
as usual. Due to the $D_{M}$ term in Eq.~(\ref{eq:density 3}),
that $\Omega_{M}=1$, $\Omega_{R}\simeq0$ and $\Omega_{\Lambda}=0$
does not imply $\Omega_{K}=0$. This does not necessarily disagree
with the WMAP result that $\Omega_{K}\approx0$~\cite{the:WMAP}
since it depends on the Universe dynamical evolution, which
is modeled differently in Eq.~(\ref{eq:density 3}).

Eq.~(\ref{eq:Friedmann}) yields the usual expression for $\mathcal{D}_L$,
the luminosity distance of a source with redshift $z$, except that $z-$dependent
density fractions $\Omega_i^{*}$ now enters $\mathcal{D}_L$:
\small
\begin{eqnarray}
\hspace{-0.3cm}
\mathcal{D}_L(z) \hspace{-0.05cm}=\hspace{-0.1cm} \frac{1+z}{H_{0}\sqrt{\Omega_{K}}} 
\sinh \hspace{-0.05cm} \bigg[  \hspace{-0.05cm} \sqrt{\Omega_{K}} \hspace{-0.1cm}
 \int_{(1+z)^{-1}}^1 
\hspace{-0.1cm}  \frac{dx}{x^{2}\sqrt{
\Omega_{K}x^{-2} \hspace{-0.05cm} + \hspace{-0.05cm} 
\Omega_{M}^{*}(z)x^{-3} \hspace{-0.05cm} + \hspace{-0.05cm}
\Omega_{R}^{*}(z)x^{-4} \hspace{-0.05cm}+ \hspace{-0.05cm} 
\Omega_{\Lambda}^{*}(z)}}\bigg] \nonumber
\end{eqnarray}
\normalsize
\vspace{-1.5cm}
\begin{eqnarray}
\label{eq: lumi dist}
\end{eqnarray}
with $x\equiv1/(1+z)$.
Likewise, the Universe age is given by:
\begin{eqnarray}
t_{0}  =\frac{1}{H_{0}}\int_{0}^{1}\frac{dx}{x\sqrt{\Omega_{K}x^{-2}+
\Omega_{M}^{*}(z)x^{-3}+\Omega_{R}^{*}(z)x^{-4}+\Omega_{\Lambda}^{*}(z)}}.\label{eq: Universe age}
\end{eqnarray}

The luminosity distance, Eq.~(\ref{eq: lumi dist}), is the quantity used to interpret the large-$z$
supernova data. We now need to model $D_M(z)$ which enters Eq.~(\ref{eq: lumi dist}) {\it {via}}
the screened density $\Omega^*_M$. (We have $\Omega^*_R = \Omega_R$, and we assume a zero
cosmological constant, $\Omega_\Lambda=0$.)

\section{Construction of $D_{M}(z)$}\label{Construction of D}

In this section, we apply our current knowledge of the evolution of large structures 
to quantitatively model the depletion function $D_M(z)$.
It is driven by:
\begin{itemize}
\item the timeline of the mass growth of large structures;
\item the masses involved in these structures relative to the total Universe mass; 
\item the geometry (mass distribution) of these structures.
\end{itemize}
The timeline and its effect on $D_{M}(z)$ is as follow: 
During the matter-radiation epoch, for $z_{eq} \lesssim 3400$, the Universe is nearly homogeneous and isotropic,
$D_{M}(z_{eq}) \approx 1$. From $15\gtrsim z\gtrsim0$, galaxies form and evolve to their present morphologies. 
At $z\simeq10$, about 10\% of the baryonic matter has coalesced into highly asymmetric 
protogalaxies~\cite{vandenBergh:2002yh,Kuutma},  with field trapping inside galaxies becoming
important. 
At $z\simeq2$, galaxies evolve to more symmetric shapes 
and the elliptical/disk galaxy ratio  increases~\cite{the:E/S ratio evolution3,the:E/S ratio evolution4,the:E/S ratio evolution5}. 
These two developments release some of the field trapped inside the galaxies~\cite{Deur DM-EPJC,Deur:2013baa}. 
During the evolution of galaxy morphologies, from $10\gtrsim z\gtrsim2$, larger structures coalesce: 
galaxies gather to form groups and protoclusters. 
At $z\simeq6$, most of the baryonic matter is in these structures, with field trapped between pairs of
galaxies. From $2\gtrsim z\gtrsim1.2$, protoclusters evolve to clusters.
For $z\lesssim1.2$, clusters start arranging themselves into more homogeneous
superstructures: filaments and sheets. This releases some of the field
trapped in between clusters. 

The different families of structure evolve with different timelines, e.g. galaxies form before groups or clusters.
Furthermore, these families have vastly different shapes and masses.  Hence, we separate $D_{M}(z)$
into a galactic part and a group/cluster/supercluster part:\footnote{The separation is approximative since the
baryonic contents of galaxies and intracluster medium interact. Also, finer separations could be considered, such as
distinguishing between group, cluster and supercluster, or between disk, elliptical and irregular galaxies. We assume 
here that separating $D_M(z)$ into two components is enough.} 
\vspace{-0.2cm}
\begin{equation}
D_{M}(z)=\xi\left[R_gD_g(z)+R_c\right]D_c(z),\label{eq:depletion model}
\vspace{-0.2cm}
\end{equation}
where $g$ stands for galaxy, and $c$ for cluster, group or supercluster. 
The normalization factor $\xi$, close to 1, accounts for possible field trapping
before galaxies start forming. 
%
$R_{i}$ (with $i=g$ or $c$) is the fraction of the baryonic mass contained in
the family $i$ at $z=0$. Such fractions should vary with $z$ but this
is factored in $D_{i}(z)$, the depletion function for the family $i$. 

Since structures grow linearly with $(1+z)^{-1}$ and since $D_i$  varies inversely to the structure mass: the larger the mass, the
smaller $D_i$, a simple choice for $D_i$ is to take $D_i \propto z$ during the growth process and $D_i$ constant
otherwise.
A better functional form is chosen in view of the following considerations: once a structure reaches a mass $m(z)$
larger than a critical mass $m_{crit}$ (that depends
on the structure geometry) at $z_{crit}$, field lines collapse, presumably quickly\footnote{This expectation is 
based the prompt transition seen in QCD between 
the string-like regime (collapsed field-lines) and the nearly free-field regime (field lines spreading 
isotropically from a charge) once the field coupling has reached a critical value~\cite{Deur:2016tte}.
In fact, the field line collapse transition in QCD has been parameterized with a FD function, see Ref.~\cite{Brodsky:2010ur}.
},
and the field gets trapped inside the structure. 
%
Hence, for a single growing system, $D_i(z)$ should
essentially be a Heaviside step-function, $D_i(z)=\big\{_{0\mbox{ for }z<z_{crit}}^{\varepsilon\mbox{ for }z\geq z_{crit}}$,
where $\varepsilon$ is the ratio of the system baryonic mass to the Universe baryonic mass. 
Since elements of a family, e.g. galaxies, reach $m_{crit}$ at different $z_{crit}$, and
since structures grow as $(1+z)^{-1}$, the overall $D_{i}(z)$ is the convolution of $m_{crit}$-weighted step-functions 
with the $m_{crit}$ probability distribution. This one is a Gaussian of width $\tau$ and centered at
average $\langle z_{crit} \rangle$ because of the initial ($z \gg z_{crit}$) normal distribution 
of mass inhomogeneities. 
If the $m_{crit}$  and $z_{crit}$ distributions are not strongly correlated, 
the resulting convolution is similar to a Fermi-Dirac (FD) function
(see Fig.~\ref{fig:D(z)_err}), which we will use to conveniently model and study $D_i(z)$.
In addition to the  FD function that encompasses the process of trapping fields into the systems,
an exponential term is added to account for a possible release of the trapped 
field, e.g.  as galaxies evolve to more symmetric morphologies or as superclusters form.
An exponential form is chosen because the release process is the reverse of the trapping process modeled by the FD
function, and $\mbox{FD}(x) \to e^x$ for large $x$. 
In all, the form for $D_i(z)$ that we will use is:
\vspace{-0.2cm}
\begin{equation}
D_{i}(z) \simeq \bigg[1-\big(1+e^{(z-z_{i0})/\tau_{i}}\big)^{-1}\bigg]+\big[A_{i}e^{-B_{i}z}\big],\label{eq:Depletion}
\vspace{-0.2cm}
\end{equation}
where the parameters $z_{i0}$, $\tau_i$, $A_i$ and $B_i$ are determined from the timeline of 
the formation of the structures of family $i$. Their interpretation is as follow:
$z_{i0} \equiv \langle z_{crit} \rangle$ is the average $z$ at which the set of structures $i$ is forming; 
$\tau_i$ is the average duration (in $z$'s scale) that such formation takes; 
$A_i$ quantifies the relative amount of structures $i$ that evolved their shapes into 
more isotropic ones; 
and $B_i$ quantifies how fast this process is.\\
First we determine these parameters for $D_g(z)$.\\
$\bullet$ {\bf Determination of $\boldsymbol{z_{g0}}$}
Approximating that galaxies grow most of their mass between $z_{g,b}=15$
and $z_{g,e}=3$ ($b$ stands for ``begin" and $e$ for ``end"), 
and evolve to their more symmetric shapes for $z<z_{g,e}$,
the FD function is centered at $z_{g0}=(z_{g,b}+z_{g,e})/2=9\pm1$.
The  $\pm1$ comes from assuming 10\% uncertainties on $z_{g,b}$ and $z_{g,e}$. \\
$\bullet$ {\bf Determination of $\boldsymbol{\tau_g}$}
The parameter $\tau_g$ characterizes
the transition width, with $\mbox{FD}(z_{0}-\tau)\approx\mbox{FD}(z_{0})/2$. 
Setting $2\tau_g=z_{g0}-z_{g,e}$,
i.e. with $\mbox{FD}(z_{g,e}) \approx 0.1$ so that the trapping
has essentially ended, yields $\tau_g=3\pm0.5$.  \\
$\bullet$ {\bf Determination of $\boldsymbol{A_g}$ and $\boldsymbol{B_g}$}
At $z<z_{g,e}$, galaxies become more symmetric. For example, at $z\simeq3$, 
the ratio of elliptical to disk galaxies 
is negligible, growing to about 50\% at $z=0$~\cite{vandenBergh:2002yh,Kuutma}.
This releases some of the field trapped in the galaxies which results in a restrengthening of gravity.  
Considering that most of the field is trapped
in disk galaxies while it is mostly released in elliptical ones, and that
the ratio of elliptical to disk galaxies is about 0.5, 
one has $D_g(z=0)\simeq0.5$, which corresponds to $A_g\simeq0.4$.
However, elliptical galaxies usually belong to clusters and the released
intragalactic field may be re-trapped between galaxy pairs. Choosing
$A_g=0.1\pm0.1$ accounts for this. It yields $D_g(z=0)\simeq0.2\pm0.1.$
Choosing $B_g=(4\pm1)z_{ge}$ makes restrengthening significant
only for $z\lesssim0.1$. 

Finally, the last galactic parameter in Eq.~(\ref{eq:depletion model}) is $R_g$, the
present baryonic mass fraction contained in galaxies. One has $R_g=0.15\pm0.10$.

We now turn to $D_c(z)$: if groups and clusters contained perfectly homogeneous and isotropically
distributed gas, extragalactic field would not be trapped inside groups or clusters. Then,
$D_c(z)$ would represent the field trapped between groups or
clusters rather than inside these structures, and one would have $R_c=1-R_g$.
To this relation, we add a term $\beta$ to account for gas anisotropy,
%
$R_c=1-R_g-\beta$.  
We assume $\beta=R_g$, i.e. that the effect 
concerns a mass similar to that which has already coalesced in galaxies. 
This is a small correction since most field trapping occurs
between the groups and clusters. 
Lastly, one needs to consider that groups and clusters are now arranging
themselves in superstructures more homogeneous than their uniformly scattered initial
distribution. This releases some of the  field trapped between groups and clusters.
Thus in all, $D_c(z)$ has the same form as $D_g(z)$, given by Eq.~(\ref{eq:Depletion}).
As for the galaxy case, the parameters for $D_c(z)$ are 
$z_{c0}\simeq(z_{c,b}+z_{c,e})/2$ and 
$\tau_c\simeq(z_{c0}-z_{c,e})/2$.
Setting $z_{c,b}=10\pm1$ (when groups/clusters start to coalesce) and $z_{c,e}=1.2$ (when superclusters start forming)
yields $z_{c0}=5.6\pm1$ and $\tau_c=2.2\pm0.5$. As for galaxies,
$B_c=4z_{g,e}=4.8\pm1.6$. $A_c$ is difficult to assess.  $A_c=0.3\pm0.15$ is tentatively chosen. 

Finally, $\xi$ in Eq.~(\ref{eq:depletion model}) accounts for
possible field trapping before galaxies started forming, i.e. for $z\gg15$, e.g. field trapping
in between the homogeneities that would latter trigger the
growth of large structures, or that which lead to the first (non-galactic) supermassive stars at $z \approx 15$. 
Since $\xi$ represents a small effect, if any, we assume $\xi=0.9\pm0.1$. 
\begin{figure}
\center
\includegraphics[width=0.45\textwidth]{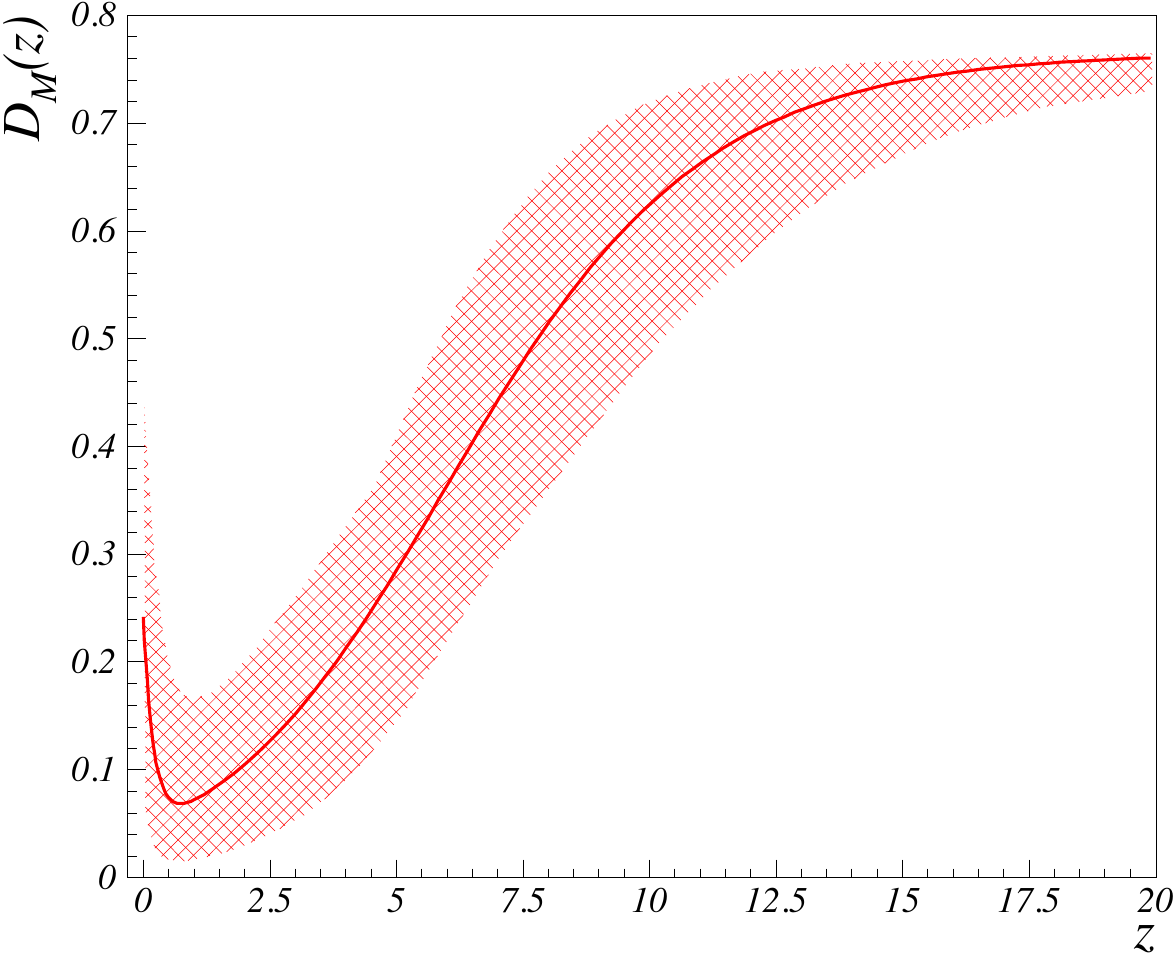}
\vspace{-0.3cm}
\caption{ Depletion factor $D_{M}(z)$, as constrained by the timeline of large structure formation 
and by the relative amount of baryonic matter pertaining to each type of structure. The width of the band 
represents the uncertainties on the value of the parameters of $D_{M}(z)$. 
The central line is obtained for the nominal values of the parameters.}
\label{fig:D(z)}
\vspace{-0.5cm}
\end{figure}

Putting together the elements of Eq.~(\ref{eq:depletion model}) produces the result shown in 
Fig.~\ref{fig:D(z)}. The $D_M(z)$ obtained for the nominal values of parameters in Eqs.~(\ref{eq:depletion model})
and (\ref{eq:Depletion}) is shown by the line.
The width of the band comes from the uncertainties on these parameters, {\it {viz}} it reflects the current state 
of our knowledge of the evolution of large structures, and of the relative amount of matter 
associated with each structure type. 
We can see the individual effects of the parameters uncertainties in Fig.~\ref{fig:D(z)_err}: it displays 
 $D_M(z)$ for the nominal values of the parameters (central line), and for the upper and lower values of one of the
 parameters while the others are kept nominal (two other lines). The total width of the 
 band is obtained by adding the effects of the uncertainties\footnote{Since the correlations between the observations used to 
 determined the values and uncertainties of the parameters are unclear, we conservatively added linearly the uncertainties 
 rather than quadratically.}. We see that the uncertainties from $\tau_c$, $z_{g0}$, $R_g$ and $\xi$ dominate.
 In the bottom right panel of Fig.~\ref{fig:D(z)_err} we show the result of using the convolution 
 of a Heaviside step-function with a Gaussian, instead of using a FD function. 
 The same nominal values of the parameters are used ($z_{i0}$ and $\tau_i$ now being, respectively, the
 center and the full width of the gaussian for family $i$). 
 Also shown is the simplest choice for $D_i(z)$: to use, instead of a FD function, 
 a function linear  between $z_{i,b}$ and $z_{e,b}$ and constant otherwise.
 The results using these different functional forms are close. 
\begin{figure}
\includegraphics[width=1.\textwidth]{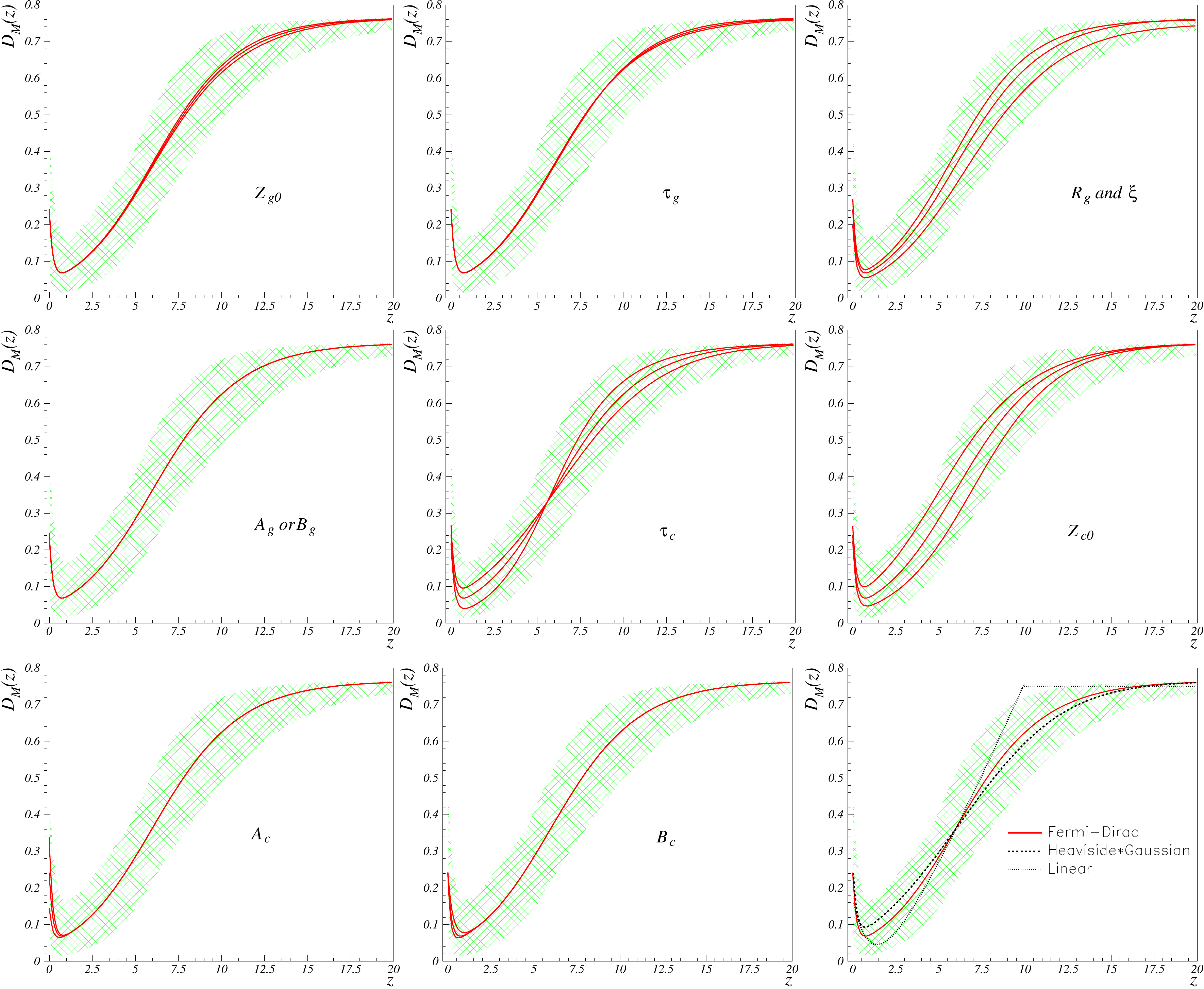}
\vspace{-0.5cm}
\caption{Individual and total contributions of the parameter uncertainties to the depletion factor $D_{M}(z)$. 
The central line shows $D_M(z)$ for the nominal values of the parameters. 
The two other lines (sometimes masked by the central line) correspond to 
upper and lower values of the parameter labelled in the panel, the other parameters being kept at their nominal values. 
The band, the same as in Fig.~\ref{fig:D(z)}, results from linearly propagating the effects of all the uncertainties.
The bottom right panel shows the results of using the convolution of a Heaviside function with a Gaussian function
(dashed line) or a simple linear function (dotted line) instead of a Fermi-Dirac function.}
\vspace{-0.5cm}
\label{fig:D(z)_err}
\end{figure}
In the next section, the factor $D(z)$ just obtained is used in the luminosity distance, Eq.~(\ref{eq: lumi dist}), 
to interpret the supernova observations without requiring $\Lambda \neq 0$ nor modifying laws of gravity or dynamics.

\section{Comparison with observations }
The compelling observations suggestive
of dark energy are: 
(1) luminosity distance 
measurements with supernovae; 
(2) the age of the Universe; 
(3) large structure formation;
(4) the cosmic microwave background (CMB);  
(5) baryon acoustic oscillations (BAO). 
We are concerned here with the foremost evidence, (1), and only
sketch how observations (2)--(5) may also be explained. Addressing them in details is
beyond the scope of a single article.

\subsection{Supernova observations}\label{sub:Supernova-observations}

Explaining supernova observations with GR's self-interaction is the focus of this article. 
These observations are that the large-$z$ ($0.1 \lesssim z \lesssim 1.5$) supernova apparent luminosities
are dimmer, {{\it {viz}} their apparent magnitudes are larger, 
than expected from a homogeneous and isotropic decelerating Universe. 
This is interpreted as evidence for an accelerating universe, i.e. for $\Lambda>0$.
However, we show in this section that lifting the approximations of homogeneity and isotropy can also explain
the observations, while keeping $\Lambda=0$.

\begin{figure}
\center
\includegraphics[width=0.5\textwidth]{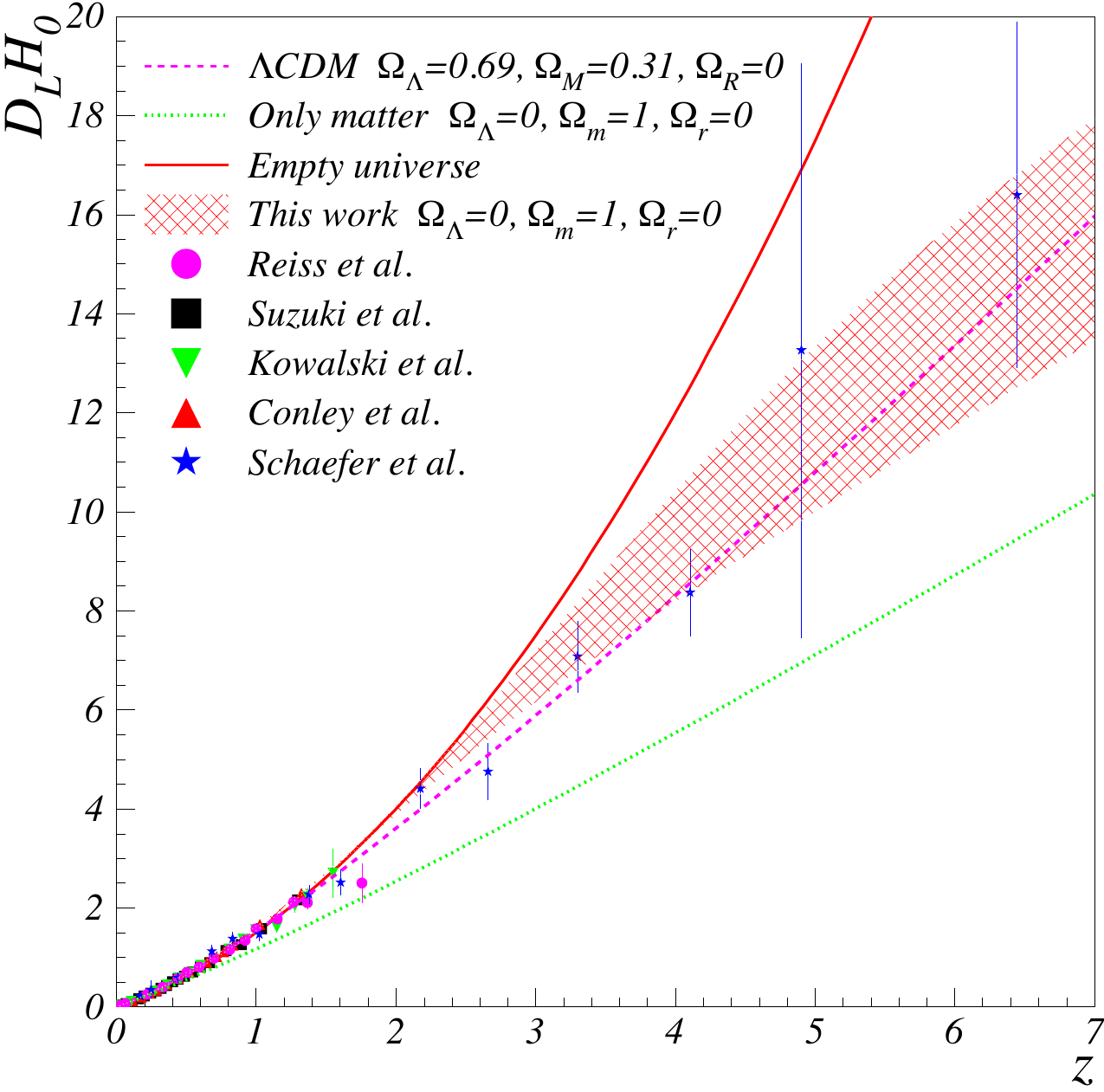}
\vspace{-0.2cm}
\caption{Apparent magnitudes of $\gamma$-ray bursts (star
symbol) and supernovae (other symbols). Larger $\mathcal{D}_L H_0$ values correspond
to fainter observed events. 
The dashed line is the expectation from the $\Lambda$CDM model. 
The dotted line is the case of a Universe with only matter 
and with the traditional approximation of homogeneity and isotropy. 
The continuous line shows the case of an empty Universe.
The band is the present work (Universe containing only baryonic matter, with gravity field
partially trapped in massive systems due to field self-interaction). It has
no free parameters adjusted to the $\gamma$-ray or  supernova data.}
\vspace{-0.5cm}
\label{fig:redshift}
\end{figure}
From the luminosity distance $\mathcal{D}_L(z)$, Eq.~(\ref{eq: lumi dist}),
the apparent magnitude $\mathcal{D}_L(z)H_0$ of events can be calculated. Assuming 
$\Lambda=0$, taking $H_0=68\pm1$ km/s/Mpc~\cite{review dark energy}, and using the depletion factor $D_m(z)$ modeled in 
Section~\ref{Construction of D} (see Fig.~\ref{fig:D(z)}), we compute the  band shown in 
Fig.~\ref{fig:redshift}. Its width stems from propagating that of $D_{M}(z)$. 
Our calculation agrees well with the $\gamma$-ray bursts~\cite{Schaefer}
and supernovae~\cite{SN data,SN data2,SN data3,SN data4}  data.
There is no adjustment to these data, all the parameters in $D_M$ being constrained by 
observations of large structure evolution.
Also shown in the figure are the calculations for the cases of a homogeneous and isotropic
Universe with only matter (dotted line), that for an empty Universe (continuous line) and the
$\Lambda$CDM (dark energy, cold dark matter) model (dashed line).

The difference between the observations and the expectation from
a homogeneous and isotropic Universe with $\Lambda=0$  is clearer by forming a 
residual apparent magnitude:
\begin{eqnarray}
r(z)=5\log\left(H_{0}\frac{\mathcal{D}_L(z)}{1+z}\right)-5\log\left(H_{0}\frac{z+z^{2}/2}{1+z}\right),
\label{eq: residual lumi dist}
\end{eqnarray}
with the last term corresponding to the empty Universe case.
Positive values of $r(z)$ indicate fainter apparent luminosities 
than expected in the case of an empty Universe. They constitute the best evidence for $\Lambda>0$.   
Our calculation of $r(z)$ with $\Lambda=0$ agrees well with the observations, see Fig.~\ref{fig:redshift residual}. 
\begin{figure}
\center
\includegraphics[width=0.5\textwidth]{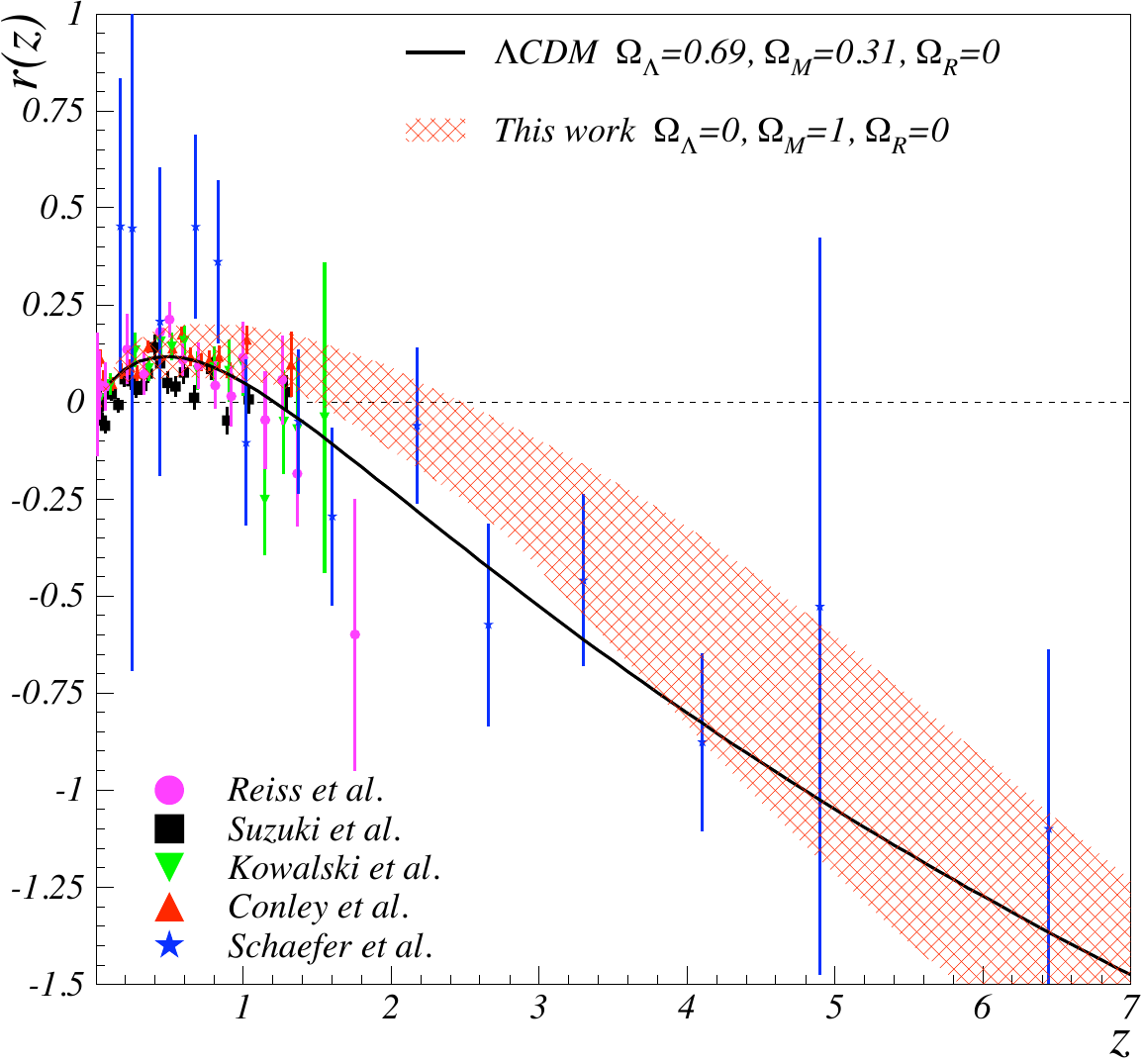}
\vspace{-0.2cm}
\caption{Residual between observed apparent magnitudes  ($\gamma$-ray bursts: star
symbol. Supernovae: other symbols) and their expectation from an empty universe. 
The continuous line is for the $\Lambda$CDM model. The band is the present
work, without any free parameters adjusted to the $\gamma$-ray or  supernova data.}
\vspace{-0.5cm}
\label{fig:redshift residual}
\end{figure}

We now outline how GR's self-interaction may also explain the observations providing less
direct evidence for $\Lambda >0$.

\subsection{Age of the Universe}
Without $\Lambda>0$, the calculated age of the Universe would be $11.7\pm0.2$ Gyr for the standard
$\Omega_{M}=0.32$ value and for $H_{0}=68\pm1$ km/s/Mpc~\cite{review dark energy}. This conflicts
with the measured age of the oldest stars, up to $\sim13.5$ Gyr.  
The $\Lambda$CDM model, with $\Omega_{\Lambda}=0.68$ and the same $H_{0}$ and $\Omega_{M}$ values,
yields $13.6\pm0.2$ Gyr. 
GR's self-interaction also solves this problem, while keeping $\Lambda=0$: Eq.~(\ref{eq: Universe age}) 
yields a compatible Universe age of $13.2\pm1.7$ Gyr.

\subsection{Large structure formation}
%
In a Universe without
gravitational self-interaction or dark matter, large structures do not have
time to coalesce. What happens in the self-interaction framework can
be sketched as follow: As $D_{M}(z)$ departs from 1, {\it{viz}} as gravity
weakens globally, energy conservation demands that the global weakening is
balanced locally by an increase of gravity within the structures themselves, thus speeding  
up their formation compared to a Universe without self-interaction. 

Since $D_{M}(z)$ evolves following the formation of large structures, 
gravity strengthens locally with the same timeline. Because 
strengthening reproduces the dynamics of galaxies and clusters~\cite{Deur DM-PLB}, 
the local effect of self-interaction is equivalent to the effect of dark matter. 
Furthermore, the position of the peak of the matter power spectrum is now given
by $k_{eq}=H_{0}\sqrt{2\Omega_{M}^{*}(0)/a_{eq}}$, with $a_{eq}$
the scale parameter at $z_{eq}$. Assuming $\Omega_{Baryon}=\Omega_{M}$
(no dark matter) and using $\Omega_{M}^{*}=\Omega_{M}D_{M}$
yield $\Omega_{M}^{*}(0)\simeq0.3$, i.e. $k_{eq}\simeq0.014$, in
agreement with observations~\cite{Tegmark:2006az}. This suggests
that the present approach is compatible with large structure formation.

\subsection{CMB and BAO}
The CMB main acoustic peak position depending on the Universe dynamical
evolution, its calculation in the present framework involves $\Omega_{M}^{*}$ rather than $\Omega_{M}$.
Thus we have now $\theta\simeq\sqrt{\Omega_{M}^{*}/z_{rec}}$ (with $z_{rec}\simeq1100$
at the recombination time), resulting in
$\theta\simeq0.8^{\circ}$, which agrees with observations~\cite{the:WMAP}.
Predicting the smaller features of the CMB and the BAO is complex and,
like for large structure formation, beyond the scope of this first article.

\subsection{Other consequence}

Field trapping naturally explains the cosmic coincidence,
i.e. that in the $\Lambda$CDM model, dark energy's repulsion currently nearly compensates
matter's attraction, while repulsion was negligible in the past and attraction
 is expected be negligible in the future~\cite{Zee}. No natural explanation exists 
 within $\Lambda$CDM for this apparently fortuitous coincidence.
 In the present approach, structure formation depletes attraction and thus, compensating it
with a repulsion, {{\it {viz}} dark energy, is unnecessary. Thus, there is no coincidence and hence no
need for explanation.

The QCD analogy to the cosmic coincidence
is that instead of accounting for the color field confinement  in hadrons, one would introduce an exotic repulsive
force to nearly counteract the strong force as it supposedly propagates outside
hadrons.

\section{Summary}
The Lagrangian of General Relativity contains field self-interaction terms that become
important for very massive systems. Their effects are unaccounted for in the studies of
galaxies and galaxy clusters since the dynamical studies of these systems rely on Newton's law of gravity. 
Accounting for field self-interaction locally strengthens gravity's binding, thereby making dark matter superfluous. 
In turn, the stronger binding in the system must be balanced by a weakening of gravity outside the system, 
as demanded by energy conservation. 
This weakening is neglected in studies of the Universe evolution because its equation is derived 
using assumptions --homogeneity and isotropy-- that suppress the effects of self-interaction, and furthermore 
disregard by definition local phenomena that could affect gravity's field, such as field trapping.

In this article, a modified Friedmann equation effectively accounting for self-interaction is derived 
from Einstein's field equation. Then, the luminosity distance formula necessary
to interpret the supernova data at large redshift $z$ is derived. Gravity's weakening is folded into
a global factor $D_{M}(z)$ that is modeled using a physically motivated function: it is constrained by the limit conditions
$D_{M}(z\gg1) \approx 1$ because of the homogeneity and isotropy of the early Universe, and 
$D_{M}(z\approx0) \ll 1$ because the growth of large structures has enabled the effects of field self-interaction.
The transition between the two limits is determined by considering that structures grow linearly with $(1+z)^{-1}$. 
The characteristic $z$ when the transition occurs, and the length of the period over which it occurs, 
are constrained by the timeline of structure formations. 
We used different functions for $D_{M}(z)$ that conform to the above constraints, and obtained similar results. 

Using the luminosity distance accounting for gravity's weakening and the modeled $D_{M}(z)$, the
large-$z$ supernova and $\gamma$-ray burst data are explained without requiring dark energy. 
No free parameters are adjusted to these data: the effect of gravity's weakening
is determined by our knowledge of large structure formation.

This approach thus explains the main observation suggestive of dark energy without requirements
beyond the standard forces and laws of physics. The basic mechanism used here is in fact well-studied: 
a similar increase of force at short range, and its consequent suppression at long range,
occurs for the strong nuclear interaction, another self-interacting force whose Lagrangian 
is similar to that of General Relativity. 
Other direct consequences of this approach are an explanation for the missing mass
in galaxies and galaxy clusters without requiring dark matter, flat rotation curves 
for disk galaxies and the Tully-Fisher relation. The direct connection between galactic missing mass
and the suppression of the Universe's deceleration eliminates the cosmic coincidence problem. 

\section*{Acknowledgments}
The author thanks  S. J. Brodsky, F. X. Girod-Gard, C. Munoz-Camacho, A. Sandorfi,
S. \v{S}irca, E. Smith, B. Terzi\'c and X. Zheng for useful discussions.


\begin{thebibliography}{00}

\bibitem{large-z 98 data} 
 A. G. Riess, {\it {et al.}}, 
 ``Observational evidence from supernovae for an accelerating universe and a cosmological constant'',
\href{http://iopscience.iop.org/article/10.1086/300499/meta}{AJ  116, 1009 (1998)}
\href{https://arxiv.org/abs/astro-ph/9805201}{[arXiv:astro-ph/9805201]} 

\bibitem{large-z 99 data}
S. Perlmutter, {\it {et al.}}, 
``Measurements of $\Omega$ and $\Lambda$ from 42 High-Redshift Supernovae,"
\href{http://iopscience.iop.org/article/10.1086/307221/meta}{ApJ 517 565 (1999)}
\href{https://arxiv.org/abs/astro-ph/9812133}{[arXiv:astro-ph/9812133]}

\bibitem{review dark energy}
C. Patrignani, {\it {et al.}}, 
 ``REVIEW OF PARTICLE PHYSICS''
\href{http://iopscience.iop.org/issue/1674-1137/40/10}{Chin. Phys. C,  40, 10000 (2016)} 
and 2017 update

\bibitem{Deur DM-PLB}
A. Deur, 
``Implications of Graviton-Graviton Interaction to Dark Matter''
\href{http://www.sciencedirect.com/science/article/pii/S0370269309004870}{PLB 676 21 (2009)}
\href{https://arxiv.org/abs/0901.4005}{[arXiv:0901.4005]}

\bibitem{Deur DM-EPJC}
A. Deur, 
``Self-interacting scalar fields at high-temperature''
\href{https://link.springer.com/article/10.1140\%2Fepjc\%2Fs10052-017-4971-x}{EPJC 77 6, 412 (2017)}
\href{https://arxiv.org/abs/1611.05515}{[arXiv:1611.05515]}

\bibitem{Zee}
A. Zee, \href{https://press.princeton.edu/titles/9227.html}{``Quantum Field Theory in a Nutshell''}, 
2003, Princeton University Press

\bibitem{Deur:2016tte} 
A. Deur, S. J. Brodsky, G. F. de Teramond, 
``The QCD Running Coupling'', 
\href{http://www.sciencedirect.com/science/article/pii/S0146641016300035}{PPNP  90, 1 (2016)}
\href{https://arxiv.org/abs/1604.08082}{[arXiv:1604.08082]};
``On the Interface between Perturbative and Nonperturbative QCD'',
\href{http://www.sciencedirect.com/science/article/pii/S037026931630048X}{PLB  757, 275 (2016)}
\href{https://arxiv.org/abs/1601.06568}{[arXiv:1601.06568]}


\bibitem{Hulse:1974eb} 
  R.~A.~Hulse and J.~H.~Taylor,
  ``Discovery of a pulsar in a binary system'',
  \href{http://adsabs.harvard.edu/doi/10.1086/181708}{Astrophys.\ J.\  {\bf 195}, L51 (1975)}


\bibitem{Deur:2013baa} 
A. Deur, 
``A relation between the dark mass of elliptical galaxies and their shape'',
\href{https://academic.oup.com/mnras/article-lookup/doi/10.1093/mnras/stt2293}{MNRAS  438, 1535 (2014)} 
 \href{https://arxiv.org/abs/1304.6932}{[arXiv:1304.6932]}

\bibitem{bullet cluster} 
 D. Clowe, {\it {et al.}}, 
``A direct empirical proof of the existence of dark matter'',
\href{http://iopscience.iop.org/article/10.1086/508162/meta}{ApJL 648 109 (2006)}
\href{https://arxiv.org/abs/astro-ph/0608407}{[arXiv:astro-ph/0608407]}


\bibitem{Tully:1977fu} 
R. B. Tully and J. R. Fisher, 
``A new method of determining distances to galaxies'',
\href{http://adsabs.harvard.edu/abs/1977A&A....54..661T}{A\&A, \textbf{54}, 661 (1977)};
S.~S.~McGaugh, J.~M.~Schombert, G.~D.~Bothun and W.~J.~G.~de Blok,
``The Baryonic Tully-Fisher relation'',
\href{https://dx.doi.org/10.1088/0004-6256/143/2/40}{Astrophys.\ J.\  {\bf 533}, L99 (2000)}
\href{https://arxiv.org/abs/1107.2934v2}{[astro-ph/0003001]}


\bibitem{the:Regge} 
T. Regge, 
``Introduction to complex orbital momenta'',
\href{https://link.springer.com/article/10.1007\%2FBF02728177}{Nuovo Cim. \textbf{14}, 951 (1959)}

\bibitem{Bali:1994de} 
  G.~S.~Bali, K.~Schilling and C.~Schlichter,
  ``Observing long color flux tubes in SU(2) lattice gauge theory'',
  \href{https://journals.aps.org/prd/abstract/10.1103/PhysRevD.51.5165}{Phys.\ Rev.\ D {\bf 51}, 5165 (1995)}
  \href{https://arxiv.org/abs/hep-lat/9409005}{[hep-lat/9409005]}


 \bibitem{Mannheim:2005bfa} 
  P.~D.~Mannheim,
  ``Alternatives to dark matter and dark energy'',
\href{https://www.sciencedirect.com/science/article/pii/S014664100500089}{Prog.\ Part.\ Nucl.\ Phys.\  {\bf 56}, 340 (2006)}
\href{https://arxiv.org/abs/astro-ph/0505266}{[astro-ph/0505266]}


\bibitem{Buchert:2007ik} 
T. Buchert,
  ``Dark Energy from Structure: A Status Report'',
  \href{https://link.springer.com/article/10.1007\%2Fs10714-007-0554-8}{Gen.\ Rel.\ Grav.\   40, 467 (2008)}
  \href{https://arxiv.org/abs/0707.2153}{[arXiv:0707.2153]}

\bibitem{Buchert:2013qma} 
T. Buchert, C. Nayet, A. Wiegand,
  ``Lagrangian theory of structure formation in relativistic cosmology II: average properties of a generic evolution model'',
  \href{https://journals.aps.org/prd/abstract/10.1103/PhysRevD.87.123503}{PRD  87 12, 123503 (2013)}
  \href{https://arxiv.org/abs/1303.6193}{[arXiv:1303.6193]}


\bibitem{the:E/S ratio evolution3}A. Dressler, {\it {et al.}}, 
``Evolution since $z = 0.5$ of the Morphology-Density relation for Clusters of Galaxies'',
\href{http://iopscience.iop.org/article/10.1086/304890/meta}{1997, ApJ 490 577}
\href{https://arxiv.org/abs/astro-ph/9707232}{[arXiv:astro-ph/9707232]}

\bibitem{the:E/S ratio evolution4}M. Postman, {\it {et al.}}, 
The Morphology - Density Relation in $z \simeq$ 1 Clusters
\href{http://iopscience.iop.org/article/10.1086/428881/meta}{2005, ApJ 623 721}
\href{https://arxiv.org/abs/astro-ph/0501224}{[arXiv:astro-ph/0501224]}

\bibitem{the:E/S ratio evolution5}O. H. Parry,  V. R. Eke, C. S. Frenk, 
``Galaxy morphology in the $\Lambda$CDM cosmology'',
\href{https://academic.oup.com/mnras/article-lookup/doi/10.1111/j.1365-2966.2009.14921.x}{MNRAS  396, 1972 (2009)} 
\href{https://arxiv.org/abs/0806.4189}{[arXiv:0806.4189]}


\bibitem{vandenBergh:2002yh}
S. van den Bergh, 
``Ten billion years of galaxy evolution'',
\href{http://iopscience.iop.org/article/10.1086/341708}{PASP114, 797 (2002)}
\href{https://arxiv.org/abs/astro-ph/0204315}{[astro-ph/0204315]}

\bibitem{Kuutma}T. Kuutma,  A. Tamm,  E. Tempel, 
``From voids to filaments: environmental transformations of galaxies in the SDSS'',
\href{https://www.aanda.org/articles/aa/abs/2017/04/aa30526-17/aa30526-17.html}{A\&A  600, L6 (2017)} 
\href{https://arxiv.org/abs/1703.04338}{[arXiv:1703.04338]}

\bibitem{Brodsky:2010ur} 
  S.~J.~Brodsky, G.~F.~de Teramond and A.~Deur,
 ``Nonperturbative QCD Coupling and its $\beta$-function from Light-Front Holography'',
  \href{https://journals.aps.org/prd/abstract/10.1103/PhysRevD.81.096010}{Phys.\ Rev.\ D {\bf 81}, 096010 (2010)}
 \href{https://arxiv.org/abs/1002.3948}{ [arXiv:1002.3948]}

\bibitem{the:WMAP} 
 G. Hinshaw, {\it {et al.}}, 
``Nine-Year Wilkinson Microwave Anisotropy Probe (WMAP) Observations: Cosmological Parameter Results'',
\href{http://iopscience.iop.org/article/10.1088/0067-0049/208/2/19/meta}{ApJS 208, 19 (2013)}
\href{https://arxiv.org/abs/1212.5226}{[arXiv:1212.5226]}

\bibitem{Schaefer}
B. E.  Schaefer, 
``The Hubble Diagram to Redshift $>6$ from 69 $\gamma$-Ray Bursts'',
\href{http://iopscience.iop.org/article/10.1086/511742/meta}{ApJ 660 16 (2007)}
\href{https://arxiv.org/abs/astro-ph/0612285}{[arXiv:astro-ph/0612285]}

\bibitem{SN data} 
G. Miknaitis, {\it {et al.}}, 
``The ESSENCE Supernova Survey: Survey Optimization, Observations, and Supernova Photometry'',
\href{http://iopscience.iop.org/article/10.1086/519986/meta;jsessionid=C44D99CF10D760C8AF8D70E0C5688A43.ip-10-40-1-105}{ApJ 666 674 (2007)}
\href{https://arxiv.org/abs/astro-ph/0701043}{[arXiv:astro-ph/0701043]}

\bibitem{SN data2}  M. Kowalski, {\it {et al.}}, 
``Improved Cosmological Constraints from New, Old and Combined Supernova Datasets'',
\href{http://iopscience.iop.org/article/10.1086/589937/meta}{ApJ 686 749 (2008)}
\href{https://arxiv.org/abs/0804.4142}{[arXiv:0804.4142]}

\bibitem{SN data3}A. Conley, {\it {et al.}}, 
``Supernova Constraints and Systematic Uncertainties from the First 3 Years of the Supernova Legacy Survey'',
\href{http://iopscience.iop.org/article/10.1088/0067-0049/192/1/1/meta}{ApJS., 192, 1 (2011)}
\href{https://arxiv.org/abs/1104.1443}{[arXiv:1104.1443]}

 \bibitem{SN data4}N. Suzuki , {\it {et al.}}, 
``The Hubble Space Telescope Cluster Supernova Survey: V. Improving the Dark Energy 
Constraints Above $z>1$ and Building an Early-Type-Hosted Supernova Sample'',
\href{http://iopscience.iop.org/article/10.1088/0004-637X/746/1/85/meta}{ApJ 746, 85 (2012)}
\href{https://arxiv.org/abs/1105.3470}{[arXiv:1105.3470]}

\bibitem{Tegmark:2006az} 
M. Tegmark, {\it {et al.}}, 
``Cosmological Constraints from the SDSS Luminous Red Galaxies'',
\href{https://journals.aps.org/prd/abstract/10.1103/PhysRevD.74.123507}{PRD  74, 123507 (2006)} 
\href{https://arxiv.org/abs/astro-ph/0608632}{[astro-ph/0608632]}

\end{thebibliography}
\end{document}